\documentclass[twocolumn,showpacs,amsmath,amssymb,showkeys,floatfix]{revtex4}

\usepackage{graphicx}
\usepackage{dcolumn}
\usepackage{bm}

\begin{document}

\title{Numerical estimation of critical parameters using the Bond entropy}

\author{Rafael~A.~Molina$^1$}
\author{Peter~Schmitteckert$^2$}
\affiliation{$^1$Instituto de Estructura de la Materia - CSIC, Serrano 123,
28006, Madrid, Spain \\
$^2$Institut f\"ur Theorie der Kondensierten Materie,
Universit\"at Karlsruhe, 76128 Karlsruhe, Germany}

\begin{abstract}
Using a model of spinless fermions in a lattice
with nearest neighbor and next-nearest neighbor interaction we show 
that the entropy of the reduced two site density matrix (the bond
entropy) can be used as an extremely accurate and easy to calculate
numerical indicator for the
critical parameters of the quantum phase transition when 
the basic ordering pattern has a two-site periodicity. The actual
behavior of the bond entropy depends on the particular characteristics
of the transition under study. For the Kosterlitz-Thouless type phase transition
from a Luttinger liquid phase to a charge density wave state the bond entropy has a local
maximum while in the transition from the Luttinger liquid to the phase separated 
state the derivative of the bond entropy  has a divergence due to the
cancelation of the third eigenvalue of the two-site reduced density matrix.
\end{abstract}


\maketitle


\section{Introduction}

A Quantum Phase Transitions (QPT) is a qualitative
change in the ground state of a quantum system as some parameter
is varied \cite{Sachdev99,Sondhi97}. Contrary to classical phase transitions,
QPTs occur at zero temperature and are due to the effect of quantum
fluctuations and not of thermal fluctuations. The previous definition is
very general, however, the abrupt change in the structure of the ground
state that define the phase transition can have different consequences
depending on the different cases. The ground state energy may become
non-analytic when approaching the critical parameter. The energy
gap between the ground state and the first excited state may go to zero 
in the critical point.
The correlations at the critical point may decay as power laws instead 
of exponentially indicating a diverging correlation length. It is
possible to find quantum systems which have some of these indications
of the QPT and not others \cite{Haldane83,Verstraete04}. For this
reason, alternative
ways for the classification of QPTs and for the numerical investigation
of the critical parameters in a QPT can be very helpful.

In recent years quantum information concepts have started to be applied
to the study of QPTs. One central concept in quantum information theory
is the concept of entanglement \cite{entanglement_review}. 
Two quantum systems in a pure state are 
entangled if their state cannot be written as the product of 
two separate pure states for each of the quantum systems.
Entanglement measures quantum correlations and as correlations are 
typically maximum at the critical points of QPTs it was realized that some 
entanglement measures may have a singularity or a maximum at the
critical point.
The amount of entanglement has been shown to be a very
sensitive quantity to the value of the critical parameter governing the
phase transition \cite{Osborne02,Osterloh02,GVidal03,JVidal04a,JVidal04b}.
In particular, concurrence \cite{Wootters} has been used to investigate
spin models and this quantity shows an extreme or singular behavior
at the corresponding critical points \cite{Osterloh02}. The block-block
entanglement between two parts of the system has also
been used, establishing connections with conformal field theory 
\cite{GVidal03,Korepin04,Keating05}.


In a recent work Gu {\it et al.} analyzed the local entanglement
and its relationship with phase transitions in the one-dimensional and
two-dimensional XXZ spin models \cite{Gu06}. The local entanglement was 
measured with the Von-Neumann entropy of the two-site density matrix. It is 
more convenient numerically than the block-block entanglement as the size of
the density matrix needed for the latter quantity  grows exponentially with
the size of the block. Using Bethe ansatz results for the one-dimensional
XXZ model Gu and coworkers showed that the local information
obtained from the entanglement entropy of the two-site density matrix is 
enough to study the phase transitions that occur in this model. Considering
blocks larger than the characteristic length scales of the system (that
in the critical point diverge) was shown to be unnecessary as the entanglement
between a block of two spins and the rest of the system is sufficient to
reveal the most important information about the system. Their results
hint to the possibility of using these properties for the numerical study
of phase transitions with small systems.

It is the purpose of this work to study numerically the local entanglement 
as a function of the size of the system. In particular, we will concentrate
in the critical points of the phase transitions. We will study a 
one-dimensional model of
spinless fermions with nearest-neighbor interactions that can be transformed
through the Jordan-Wigner transformation to the
XXZ model with spins at each site $S_j=1/2$ \cite{JordanWigner}. 
For repulsive interaction $V_1=2.0$ the ground state 
performs a Kosterlitz-Thouless type phase transition
from a Luttinger Liquid to a charge density wave, which corresponds
to an antiferromagnetically ordered state in the spin picture. 
For attractive interaction $V_1=-2.0$ there is a phase transition 
to a phase separated state, 
which corresponds to a ferromagnet in the spin picture.
We will show that the bond entropy $S_\text{bond}$ enables us to determine 
the critical points of both quantum phase transitions with an astonishing accuracy,
albeit they present very different characteristics and symmetries.
We will show how these differences are reflected in the behavior of the
bond entropy. 
Finally we include next-nearest-neighbor interaction  to test 
the generality of our finding.
The effect of longer ranged interaction was studied before in the context of
multiple umklapp scattering \cite{Schmitteckert04}.
In this case the interaction can now lead to phase transitions at fillings
different from 1/2 and to ordering patterns with increased unit cell. 

\section{Bond-entropy for the nearest-neighbor interaction model}

We shall consider the one-dimensional spinless fermions model with
next-neighbor interactions.
\begin{equation}
\label{eq:H_spinless}
\hat{H}=-t \sum_{i=1}^{L} \left( \hat{c}_i^{\dagger} 
\hat{c}_{i-1}^{\phantom{\dagger}} + 
\hat{c}_{i-1}^{\dagger} \hat{c}_{i}^{\phantom{\dagger}}\right) 
+ \sum_{i=1}^{L} 
V_1 (\hat{n}_i-\frac{1}{2}) (\hat{n}_{i-1}- \frac{1}{2}),
\end{equation}
where the operators appearing in the formula are the usual fermionic creation, 
annihilation, and number operator at site i, 
$L$ is the total number of sites, $t$ is the hopping matrix element between
neighboring sites, and $V_1$ is the nearest-neighbor interaction strength. 
An important property of this model
is that it can be transformed into the XXZ spin $S=1/2$ model through the 
Jordan-Wigner transformation
\cite{JordanWigner,Nagaosa}. 
\begin{eqnarray}
\label{eq:JordanWigner}
\hat{S}_j^-  = & \exp{\left(-i\pi\sum_{\ell=1}^{i-1}\hat{c}_{\ell}^{\dagger}
\hat{c}_{\ell}^{\phantom{\dagger}}\right)} c_j, \\
\hat{S}_j^z  = & \hat{n}_j-1/2.
\end{eqnarray}
For an even number  of particles a phase term 
appears in the boundary condtion 
when we apply the Jordan-Wigner transformation to Hamiltonian
(\ref{eq:H_spinless}). As we are not interested in this
even-odd effect we will consider
periodic (anti-periodic) boundary conditions, $c_0 \equiv c_M$ ($c_0 \equiv -c_M$), 
for N odd (even) and both models will be equivalent in our examples.
Although we will mainly consider the spinless fermion model 
we will make comments
regarding the equivalent behavior of both models when we believe it will
be useful to clarify some situation (specially in the ``ferromagnetic'' phase).
The Hamiltonian (\ref{eq:H_spinless}) commutes with the 
total $\hat{N}=\sum \hat{n}_i$ operator
so the total number of particles is a conserved quantity and we will
consider subspaces with a definite number of fermions, equivalent to
consider subspaces with a definite value of $S_z$ in the XXZ model.

This model has an interesting phase diagram depending on the value of the
ratio of the interaction parameter and the hopping term 
$V_1/t$ and also on the number of particles $N$. 
Without loss of generality we can consider $t=1$ and consider the phases
as we change $V_1$. 
For $V_1<-2$ the equivalent spin system
is ferromagnetic and the ground state is fully spin polarized. When we
cross the first transition point $V_{ca}=-2$ the ground state of the
system can be shown to be non-degenerate and with spin $S=0$ 
\cite{Lieb62}. Only in the half-filled case, $N=L/2$, there is another 
transition point at $V_{cb}=2$. 
For $V_1>2$ the system is a charge density wave type insulator, 
the transition is of the
Kosterlitz-Thouless type and the order parameter depends
exponentially on the difference $V-V_{cb}$ 
making an accurate numerical determination of the
transition point notoriously hard.

We will define the bond-entropy as the Von Neumann entropy of the reduced
density matrix of two-neighboring sites $\hat{\rho}_{i i+1}$. As a result
of the conservation of $N$ the reduced density matrix can be written as
a $4 \times 4$ matrix with three sectors of $N=0$, $N=1$, and $N=2$. In
the two-site basis $\left|00\right>$, $\left|01\right>$, $\left|10\right>$,
$\left|11\right>$ it can be represented as, 
\begin{equation}
\label{eq:general_rho}
\hat{\rho}_{i i+1}= \left( \begin{array}{cccc}
u^{-} & 0 & 0 & 0 \\
0 & \omega & z & 0 \\
0 & z^{*} & \omega & 0 \\
0 & 0 & 0 & u^{+}
\end{array} \right).
\end{equation}

For this particular model using its invariance under translations it can
be shown that 
we can write this matrix elements in terms of certain correlation functions
\cite{Wang02,Gu05},
\begin{equation}
\label{eq:um}
u^{-}=1+\left<\hat{n}_i \hat{n}_{i+1}\right>-\frac{3}{2}\left<\hat{n}_i\right>
-\frac{1}{2}\left<\hat{n}_{i+1}\right>.
\end{equation}
\begin{equation}
\label{eq:up}
u^{+}=\left<\hat{n}_i \hat{n}_{i+1}\right>+\frac{1}{2}\left<\hat{n}_i\right>
-\frac{1}{2}\left<\hat{n}_{i+1}\right>.
\end{equation}
\begin{equation}
\label{eq:omega}
\omega = \frac{1}{2}\left<\hat{n}_{i}\right>+
\frac{1}{2}\left<\hat{n}_{i+1}\right>-
\left<\hat{n}_i \hat{n}_{i+1}\right>.
\end{equation}
\begin{equation}
\label{eq:z}
z = \frac{1}{2}\left(\left<\hat{c}_i^{\dagger} \hat{c}_{i+1}^{\phantom{\dagger }}\right>+ \left<\hat{c}_{i+1}^{\dagger} \hat{c}_{i}^{\phantom{\dagger}}\right>\right).
\end{equation}

We define the bond entropy $S_\text{bond}^i$ as the Von Neumann entropy of
the two-site density matrix $\hat{\rho}_{i\,i+1}$. In the model under study,
the invariance under translations also implies that $S_\text{bond}^i$ does
not depend on the site $i$.
\begin{equation}
\label{eq:sbond}
S_\text{bond}= -\sum_{j=1}^{4} \lambda_j \ln \lambda_j,
\end{equation}
where $\lambda_j$ are the four eigenvalues of the reduced two-site 
density matrix $\hat{\rho}_{i\,i+1}$.

\section{Numerical results}

In this section we will show the numerical results for the bond entropy 
as a function of the size of the system using
the DMRG algorithm \cite{White92}. 

In Figure 1 we show the results of $S_\text{bond}$ at half filling 
for different number of sites $L$. The behavior of the bond entropy around the
two critical points is very different, reflecting the different changes
in the symmetries and correlations of the ground state. 
The slope of $S_\text{bond}$ diverges around $V_{ca}=-2$ 
(we will explain that in more detail in the next section), 
while $S_\text{bond}$ is continuous but has a local maximum in the proximity of
the second critical point $V_{cb}=2$. In both cases one can understand
the behavior of $S_\text{bond}$ from the behavior of the correlation functions
in the different phases \cite{Gu06}. More importantly, one can estimate
with extraordinary precision the value of the critical parameter from
very small system sizes even in the case of the  transition to the CDW phase.
In addition, $S_\text{bond}$ has a minimum for $V_1=0$.

\begin{figure}
\begin{center}
\includegraphics[width=8cm,height=5cm]{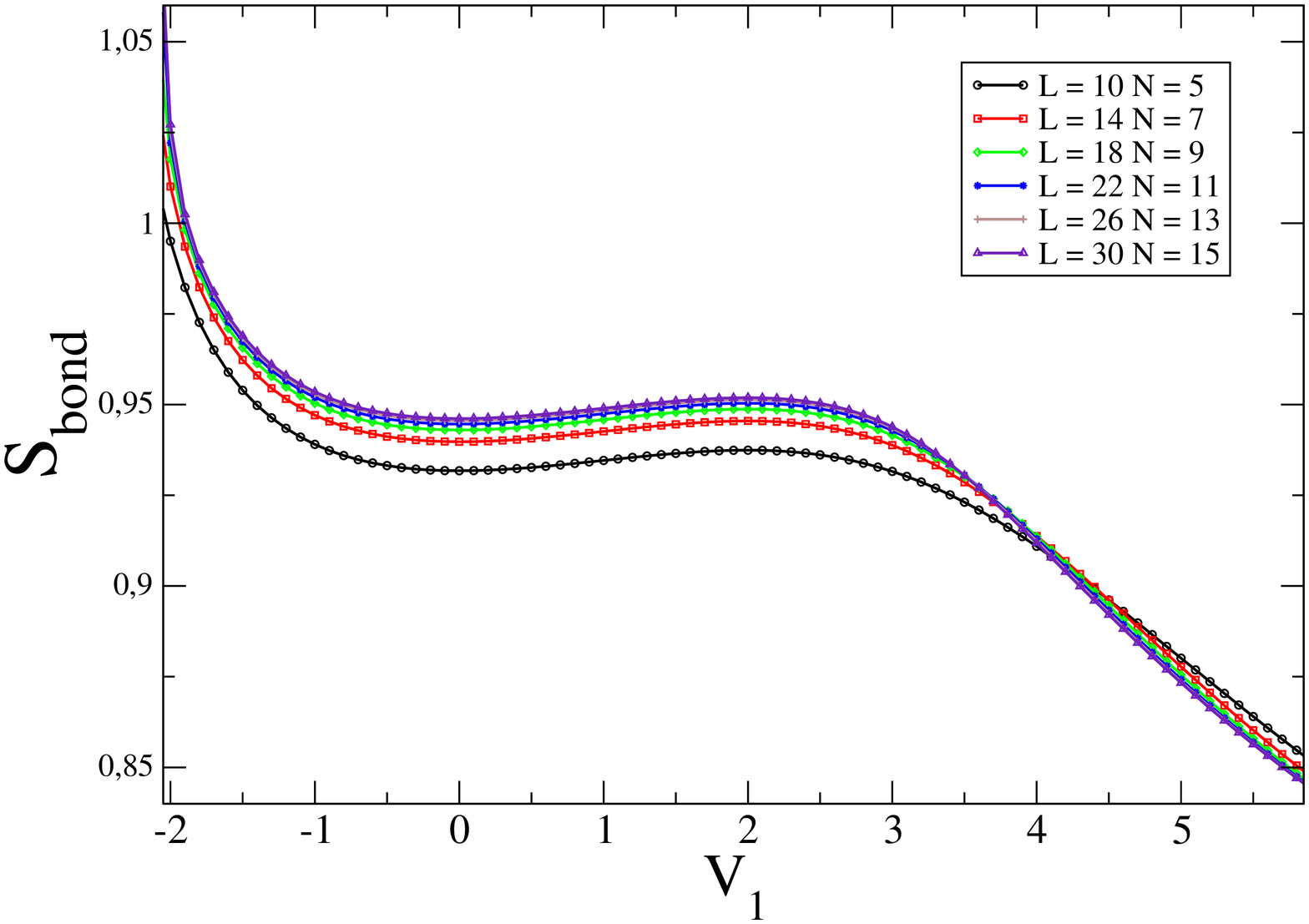}
\caption{\label{fig:sbondhf} (Color online) Bond entropy $S_\text{bond}$ as
a function of the interaction for different sizes of the spinless fermions
ring $L$ at half filling $N=L/2$. The results for $L=26$ and $L=30$ are
hardly distinguishable. We have a maximum of $S_\text{bond}$ at $V_1=2$ with
extremely good approximation, see next figure, that marks the CDW insulator
transition. The slope of the entropy diverges at $V_1 = -2$ marking the appearance
of the ferromagnetic transition.}
\end{center}
\end{figure}

In Figure 2 we see a zoom of the last figure 1 for one particular case with
parameter $L=30$, $N=15$ in the region around $V_1=2$. When we calculate 
numerically the position of the maximum with DMRG we obtain the value
of $V_1$ with a precision better than $10^{-5}$. Of course, in order
to take full advantage of these properties of $S_\text{bond}$ we need a
very accurate algorithm such as DMRG. We used at least 500 states per block
for the $L \le 30$ sites leading to an discarded entropy below $10^{-9}$, 
and 1400 states per block for the 36 site system (see below),
including the five lowest lying states in order to treat the degeneracies correctly,
leading to a discarded entropy typically below $10^{-10}$, and up to  $2\cdot 10^{-8}$ close
to the phase transitions and applied always eleven finite lattice sweeps.
Close to the CDW-I -- Luttinger liquid transition (see below)
we used at least 2050 states per block and only two low lying states to check our results.
This resulted in a discarded entropy below $10^{-12}$.
We'd like to note that despite the large number of states the DMRG runs are much cheaper
as compared to the calculations in  \cite{Schmitteckert04}, since no resolvent has to be computed,
e.g.\ the largest run took about a hundred CPU minutes.

\begin{figure}
\begin{center}
\includegraphics[width=8cm,height=5cm]{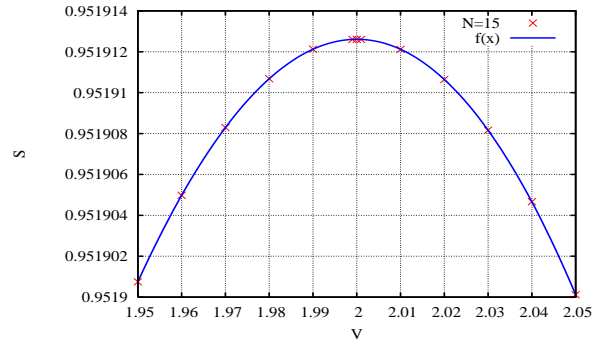}
\caption{\label{fig:s3} (Color online) Behavior of $S_\text{bond}$
as a function of $V_1$ in the neighborhood of the phase transition
at $V_1=2$ in the case $L=30$, $N=15$. We can numerically pinpoint the
maximum of the curve at $V_1=2$ with a precision better than $10^{-5}$.
The function $f(x)$ is the second order polynomial fit used to obtain the
maximum of the curve.
}
\end{center}
\end{figure}

In Figure 3 we see some examples comparing results at half-filling and
outside half-filling. The qualitative behavior of the bond entropy is
exactly the same around the first critical point $V_{ca}=-2$, but the maximum
of the bond entropy around $V_{cb}=2$ disappears as soon as we move outside
half-filling reflecting the absence of the CDW transition 
for fillings different from $1/2$.

\begin{figure}
\begin{center}
\includegraphics[width=8cm,height=5cm]{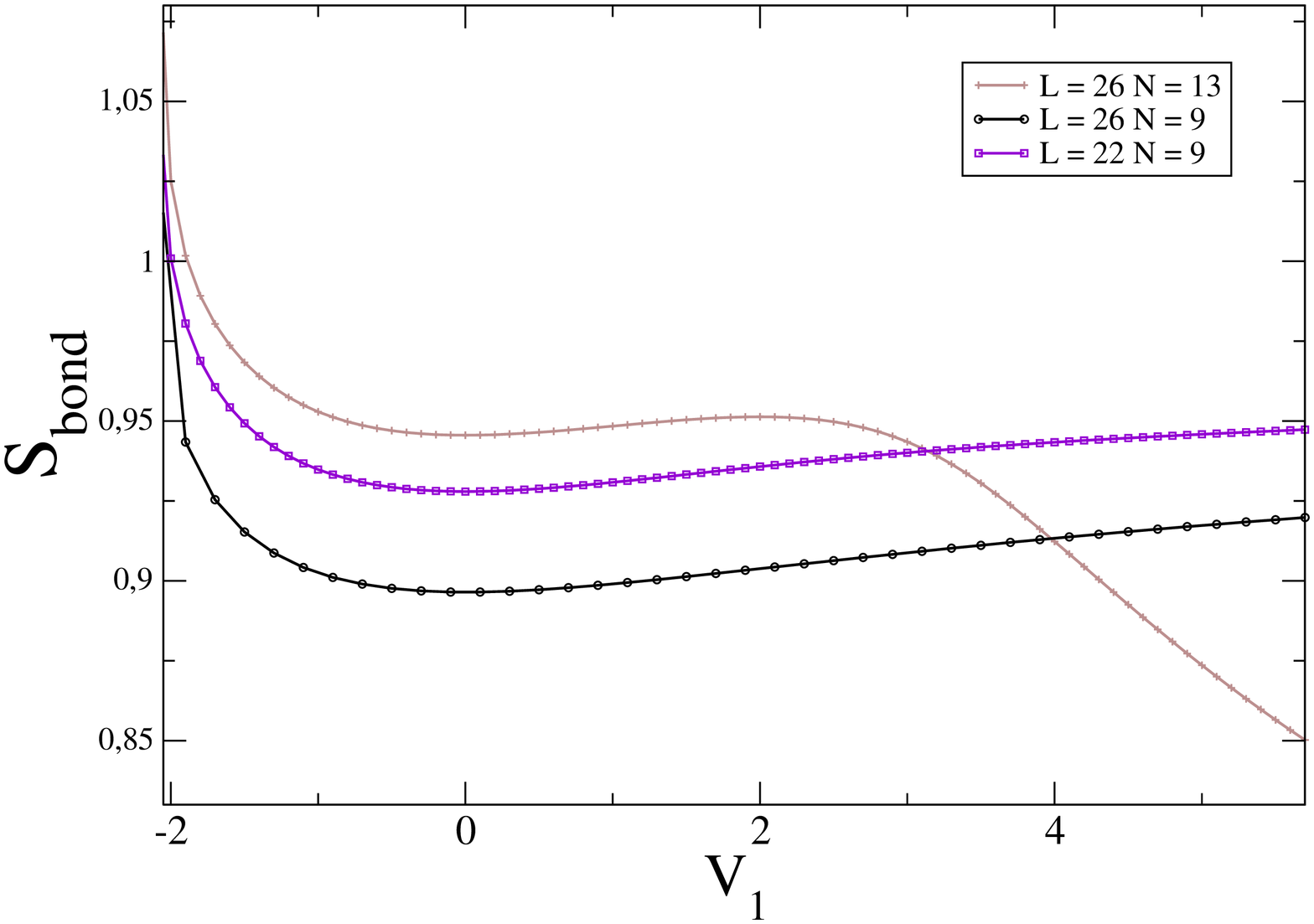}
\caption{\label{fig:f3} (Color online) Behavior of $S_\text{bond}$
as a function of $V$ for $L=26$, $N=13$ (half-filling) and
$L=26$, $N=9$, and $L=22$, $N=9$ (outside half-filling). We observe
the same behavior in the ferromagnetic transition around $V_{ca}=-2$
but a complete different one in the transition around $V_{cb}=2$.
}
\end{center}
\end{figure}

\section{Ferromagnetism and the two-site density matrix}

\begin{figure}
\begin{center}
\includegraphics[width=8cm,height=5cm]{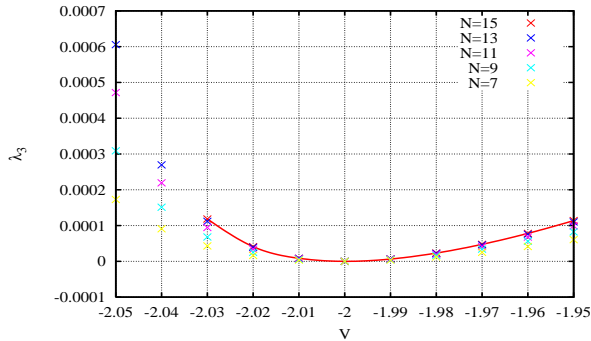}
\caption{\label{fig:lambda3} (Color online) Behavior of the third
eigenvalue of the two-site density matrix $\lambda_3$
as a function of $V$ in the neighborhood of the phase transition
at $V_1=-2$ for different sizes at half filling. 
$\lambda_3$ vanishes at $V_1=-2$ with very high precision.
}
\end{center}
\end{figure}

As we have mentioned before the slope of $S_\text{bond}$ diverges at $V_1=V_{ca}$. 
In Fig. \ref{fig:lambda3} we show the results for the value of the third
eigenvalue of the two-site density matrix $\lambda_3$ for different sizes.
We can reach a very high numerical precision in the determination
of the ferromagnetic critical point studying the cancellation of the
third eigenvalue of the two-site density matrix.

In the thermodynamic limit at $V_1=-2$, $\omega= z = 1/4$ in Eq. 
(\ref{eq:general_rho}). One can immediately see that the equality of both 
matrix elements implies that one of the eigenvalues of the density matrix 
is zero, which leads to the singularity of  $S_\text{bond}$.
If one tries a direct numerical examination of the values of the 
correlation functions independently one does not get a very accurate
estimation of the critical parameter. However, the examination of the
particular combination appearing in the Von Neumann entropy of the two-site
density matrix allows a very accurate calculation even with very small
system sizes due to the fact that although $\omega$ and $z$ converge
slowly to the thermodynamic limit value $1/4$, their difference converges
very quickly to zero at the critical value of the interaction.

\section{Next nearest-neighbor interaction model}

In order to test the generality of our conclusions and to obtain a critical
parameter of a phase transition in a model not solvable with Bethe ansatz
we add next-nearest neighbors interaction,
\begin{eqnarray}
\label{eq:H_nnn}
\hat{H} & = &-t \sum_{i=1}^{L} \left( \hat{c}_i^{\dagger} 
\hat{c}_{i-1}^{\phantom{\dagger}} + 
\hat{c}_{i-1}^{\dagger} \hat{c}_{i}^{\phantom{\dagger}}\right) \nonumber \\
& +& \sum_{i=1}^{L} 
V \left(\hat{n}_i-\frac{1}{2}\right) \left(\hat{n}_{i-1}-\frac{1}{2}\right) 
\nonumber \\
& + & \sum_{i=1}^{L} 
V_2 \left(\hat{n}_i-\frac{1}{2}\right) \left(\hat{n}_{i-2}-\frac{1}{2}\right) ,
\end{eqnarray} 
where $V_2$ is the strength of the interaction between sites separated
by two lattice spacings. This Hamiltonian has been used to study
the physics of materials that exhibit multiple phase transitions. Usually one
considers $V_2 < V_1$ as one expects the interaction to reduce with distance.
However, there can be exceptions if the nearest-neighbor
interaction is supressed by the lattice geometry.

\begin{figure}
\begin{center}
\includegraphics[width=8cm,height=5cm]{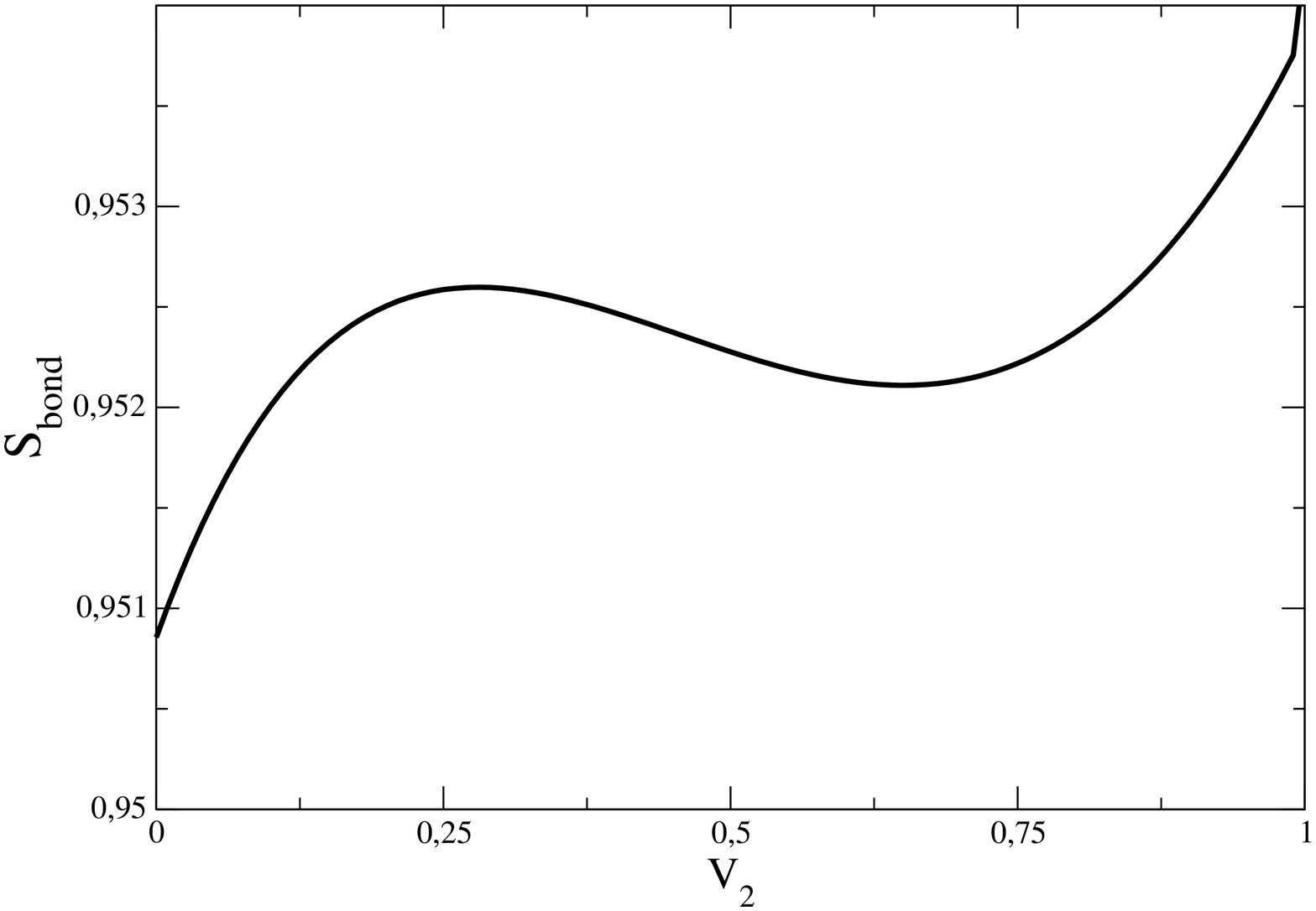}
\caption{\label{fig:V2} 
Bond entropy as a function
of $V_2$ in the line $V_2=5-2V_1$ for $L=36$
and $N=18$. The maximum can be
used to estimate the critical point at $V_{2,c}=0.280$ with very
high precision.
}
\end{center}
\end{figure}

\begin{figure}
\begin{center}
\includegraphics[width=8cm,height=5cm]{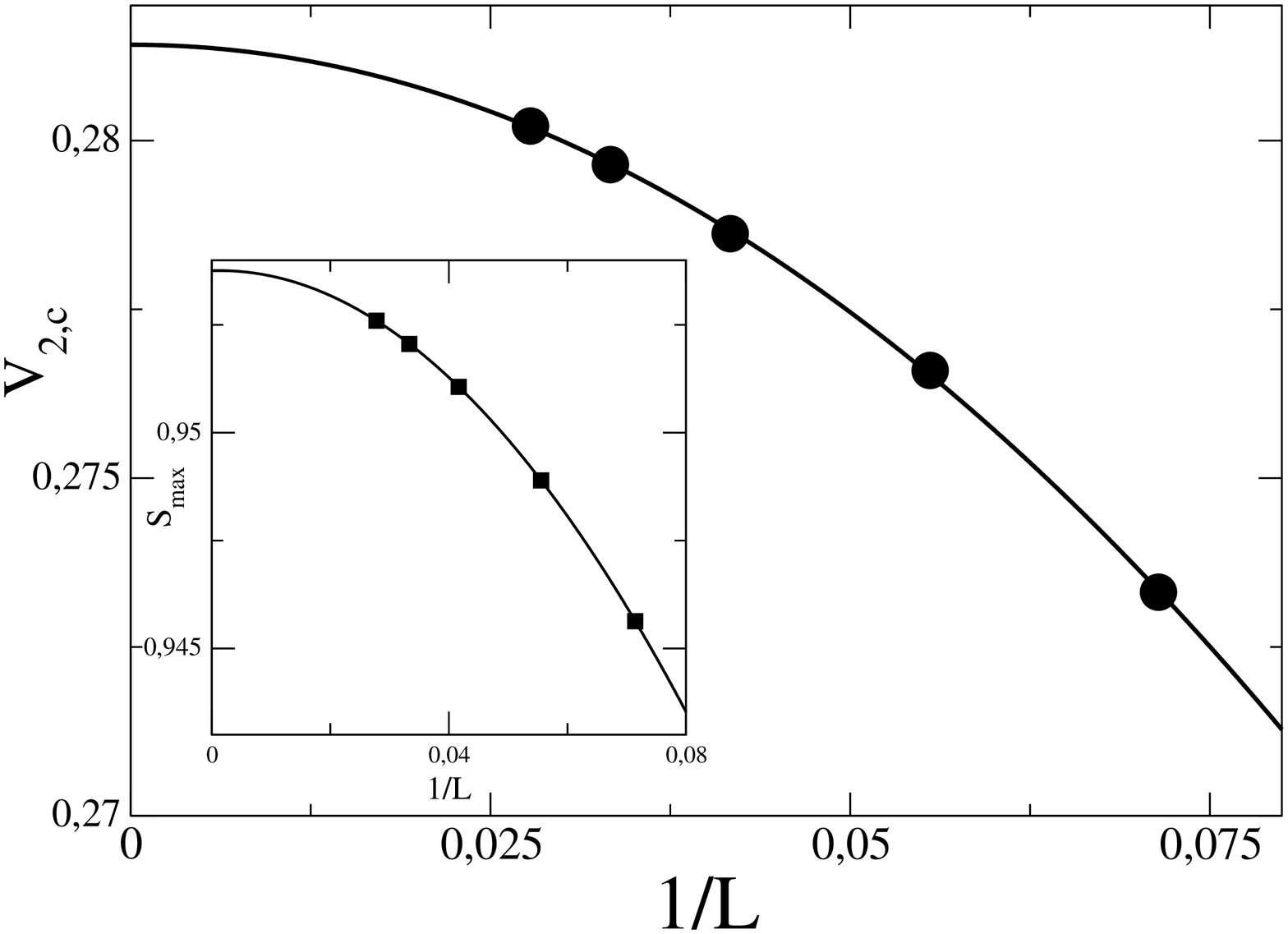}
\caption{\label{fig:Size_scaling} Finite size scaling for the position of
the maximum of the bond entropy $V_{2,c}$ as a function of $1/L$. 
The continuous line is a second order polynomial fit.
It is used to extrapolate the value
for $L=\infty$. In the inset we show the value of the local maximum
of $S_\text{bond}$ as a function of $1/L$, a second order polynomial also
fits very well the numerical results.
}
\end{center}
\end{figure}

The phase diagram of the model represented by the Hamiltonian 
(\ref{eq:H_nnn}) has been studied as a function of $V_1$ and $V_2$ by
Schmitteckert and Werner \cite{Schmitteckert04}. In this paper the authors
used DMRG to calculate the ground state curvature. The phase diagram depends
on the filling, commensurability effects are extremely important due to
the multiple umklapp scattering. If we concentrate on half-filling and
repulsive interactions we have
a charge density wave (CDW) phase in which the ground
state is twofold degenerate with ordering pattern  
($\bullet$$\circ$$\bullet$$\circ$) and
($\circ$$\bullet$$\circ$$\bullet$). Here $\circ$ denotes a vacant and
$\bullet$ denotes an occupied site. In phase CDW II the ground state is
fourfold degenerate with ordering pattern
($\bullet$$\bullet$$\circ$$\circ$),
($\circ$$\bullet$$\bullet$$\circ$),
($\circ$$\circ$$\bullet$$\bullet$), and
($\bullet$$\circ$$\circ$$\bullet$). We will follow reference 
\cite{Schmitteckert04} and study the critical parameters along the line
$V_2=5-2V$. For example, studying systems of sizes up to $L=60$ they
obtained a critical point for the transition between the CDW I phase and
the Luttinger Liquid phase as 
$(V_{1,\rm c},V_{2,\rm c})=(2.4\pm0.05,0.2\pm0.1)\}$. In Fig. (\ref{fig:V2})
we show results for the ground state bond entropy as a function of $V_2$
for along the previous mentioned line for $L=36$ and $N=18$. Even from the 
small size used we can accurately determine the critical $V_2$
as $V_{2,\rm c} = 0.280$, which is within the error bars previously 
given by Schmitteckert and Werner \cite{Schmitteckert04}. 
The determination of the critical parameter
for the transition between CDW I and the Luttinger liquid is done with much
less numerical work as compared to the finite size analysis of excitation gaps
and the ground state curvate in \cite{Schmitteckert04}.
Notably, the finite size corrections are smaller,
e.g.\ the results of the same analysis with $L=18$ 
already gives a critical parameter of $V_{2,\rm c} = 0.277$.
In Fig. \ref{fig:Size_scaling} we show the numerical results for the value 
of the next-neighbor interaction $V_2$ in which we have a
local maximum of $S_\text{bond}$ ( along the same line as before) as a 
function of the inverse of the total length of the system $1/L$. We have 
calculated numerically the bond entropy at each size with an interval of 
$0.001$ in $V_2$ in the region around the maximum of $S_\text{bond}$, except 
in the case of $L=36$ where we have used an interval of $=0.0002$. The values 
of $V_2$ in the maximum where obtained through a second-order polynomial fit 
of the numerical results for $S_\text{bond}$. The actual value of the interval 
used was not very critical as the fits were very good. With another 
second-order polynomial fit we can extrapolate the calculated values to obtain 
the result for the thermodynamic limit $V_{2,c}=0.2814 \pm 0.0001$. In the 
inset of the figure we can see the numerical results used in the extrapolation 
of the value of the maximum of the bond entropy in the critical point. 
The extrapolated value being $S_\text{max}=0.95385 \pm 0.00001$.

In this case the bond entropy and the two-site density matrix does not
give very useful information about the phase transitions to charge
density wave phases with ordering patterns with basic sizes bigger than
two. We obtain no clear signature in the bond entropy for the
quantum phase transition to CDW II. One may have to study  entropies of
density matrices of blocks with at least the size of the basic ordered
block of the phases we are looking at. Also, one could try to study the
bond entropy for the excited states.
The numerical determination of critical parameters in this case is out
of the scope of this study of the bond entropy and will be subject of
future work. We note that in the case of CDW II the ground state is fourfold
degenerate in the thermodynamic limit. 
However, the degeneracy of the two lowest lying
states is lifted by finite size effects.

\section{Conclusions}

The bond entropy defined as the Von Neumann entropy of the two-site density
matrix can be a very effective tool for the study of phase transitions and
critical parameters. Its behavior depends on the correlations in the
ground state of the system.

We have studied the bond entropy for a model of spinless fermions with
nearest-neighbor interactions and periodic boundary conditions. The size
dependence of its behavior near the two critical points in the model has
been studied in detail, showing an amazing precision in the estimation
of the critical parameter. We  have also studied a model with next-nearest
neighbor interactions. We could determine the critical point
of the phase transition from the Luttinger liquid to the CDW I state with
an ordering pattern of period two. 
If the fundamental block contains only
two sites we show that the bond entropy displays a clear signature of the
quantum phase transitons
and allows for the determination of the critical parameters. 
The bond entropy of the
ground state could not be used for the transition to CDW II with ordering 
pattern of period four.  In this case we may have to turn to a block entropy
of higher size. In general, we can say that the bond entropy can be used
as a numerical indicator for phase transitions but the actual behavior
of the bond entropy is not universal and will depend on the QPT under
study. Our results should open the way to the 
numerical study of phase transitions with small sized systems.  


\acknowledgments

RAM wishes to acknowledge useful discussions with J. Dukelsky. He also
acknowledges finantial support at the 
Instituto de Estructura de la Materia-CSIC by an I3P contract funded by
the European Social Fund. This work is supported in part by
Spanish Government Grant FIS2006-12783-C03-01.

\end{document}